\documentclass[acmtog, authorversion]{acmart}
\AtBeginDocument{%
  \providecommand\BibTeX{{%
    \normalfont B\kern-0.5em{\scshape i\kern-0.25em b}\kern-0.8em\TeX}}}

\setcopyright{acmcopyright}
\copyrightyear{2024}
\acmYear{2024}

\acmConference[AIMLSystems 2024]{The Third International Conference on AI-ML Systems }{October 8-11,
  2024}{Lousiana, USA}
\acmSubmissionID{120}

\begin{document}

\title{CultureVo: The Serious Game of Utilizing Gen AI for Enhancing Cultural Intelligence}

\author{Ajita Agarwala}
\email{ajita.agarwal.4dec@gmail.com}
\affiliation{%
   \institution{ \url{CultureVo, Inc.} 
     \city{Delhi}
     \country{India}}
}

\author{Anupam Purwar}
\orcid{0000-0002-5580-4917}
\email{anupam.aiml@gmail.com}
\affiliation{%
  \institution{Independent}
  \city{Delhi}
  \country{India}}

\author{Viswanadhasai Rao}
\orcid{0000-0002-5580-4917}
\email{viswanadhasai_nissankararao@srmap.edu.in}
\affiliation{%
  \institution{SRM University, Guntur}
  \city{Andhra Pradesh}
  \country{India}}

\renewcommand{\shortauthors}{Ajita, et al.}


\begin{abstract}
CultureVo, Inc. has developed the Integrated Culture Learning Suite (ICLS) to deliver foundational knowledge of world cultures through a combination of interactive lessons and gamified experiences. This paper explores how Generative AI powered by open source Large Langauge Models are utilized within the ICLS to enhance cultural intelligence. The suite employs Generative AI techniques to automate the assessment of learner knowledge, analyze behavioral patterns, and manage interactions with non-player characters using real time learner assessment. Additionally, ICLS  provides contextual hint and recommend course content by assessing learner proficiency, while Generative AI facilitates the automated creation and validation of educational content.
\end{abstract}

\ccsdesc[500]{Computing methodologies}
\ccsdesc[500]{Computing methodologies~Natural language generation}
\keywords{ Large Language model (LLM), Serious Game, Pedagogical agents, Cultural intelligence, Integrated Culture Learning Suite  (ICLS)}


\maketitle
\section{Introduction}
The rise of globalization and technology has significantly shifted individual responsibility, extending it from personal actions to the well-being of people worldwide \cite{Globalcitizenship}. In this context, cultural intelligence has become increasingly essential, as it equips individuals to navigate and engage effectively within our diverse and interconnected global community \cite{GlobalLiteracy}. It is crucial for fostering effective communication, mutual respect and successful collaboration across diverse cultural contexts \cite{CulturalIntelligence}. Developing this intelligence is often impeded by information overload, fragmented resources, and limited exposure to a variety of cultures. Traditional educational methods may struggle with engagement, and physical immersion experiences are frequently limited by cost and logistical constraints. This makes innovative solutions essential for promoting cultural intelligence, which can help reinforce the understanding that our differences can positively impact our ability to work together, perform better, feel better, and ultimately build a better world. \cite{ManagementResearch}.

CultureVo, a serious game, effectively addresses these challenges through the Integrated Culture Learning Suite (ICLS) by offering a gamified, interactive learning platform that significantly enhances cultural intelligence. Through its curated, precise content tailored to individual learning needs, CultureVo overcomes the limitations of traditional search engines by delivering focused, relevant information. The platform boosts user engagement and motivation with personalized access and interactive features while addressing privacy concerns through advanced knowledge-augmented retrieval techniques. Moreover, CultureVo provides accessible digital cultural immersion, integrates fragmented resources, and ensures secure data management \cite{Virtualreality}. By incorporating web interface, cloud computing, and large language models, CultureVo presents a comprehensive and user-friendly solution for effective cultural learning. This paper outlines the Integrated Culture Learning Suite (ICLS) developed by CultureVo as a serious game prototype, emphasizing its impact on advancing cultural intelligence and enhancing learner engagement.
\section{System Overview}
Each ICLS training module integrates three fundamental components to enhance cultural competence systematically. The Cultural Treasury functions as an extensive repository of reference material, offering in-depth coverage of the cultural aspects of diverse countries. This component includes detailed glossaries, thematic summaries, and analytical frameworks that elucidate the terminology and cultural themes discussed in the lessons. It is designed to support advanced research and provide contextual understanding, serving as a comprehensive resource for learners seeking to deepen their cultural insights. World Wise is a sophisticated assessment tool designed to consolidate and evaluate learners’ grasp of the concepts introduced in the training modules. It features a quiz format that not only tests knowledge but also allows for iterative improvement and mastery of cross-cultural concepts. The Culture Scribe is an advanced interactive system designed to enhance and expand the user's cultural knowledge. It employs sophisticated analytics to assess user engagement by monitoring the time spent on country-specific content, quiz attempts, and quiz performance. This data is used to calculate a proficiency score for each user. The system then tailors interactions based on the user's unique learning needs, ensuring that each engagement effectively addresses gaps in their understanding of the global cultural landscape. Together, these components create a robust educational framework, combining detailed reference, evaluative assessment, and interactive practice to foster a thorough and scientific approach to cultural intelligence development. \cite{MacNab2012-lw}

Each component of ICLS is meticulously designed to enhance the user’s cultural intelligence through practical and interactive methods, embodying the principles of a serious game. The Cultural Treasury functions as a thorough resource, delivering detailed information and insights into diverse cultures. For instance, if you need to understand Japanese business etiquette, this resource provides detailed explanations and key cultural concepts, laying the groundwork for informed and respectful interactions.

World Wise integrates a dynamic quiz format that challenges users to test the cultural knowledge they’ve gained from the Cultural Treasury. By engaging in quizzes that simulate intercultural comparisons, such as exploring historical underpinnings in different social contexts or analyzing cultural practices across various societies, users can apply and reinforce their understanding. This interactive approach not only assesses their grasp of key concepts but also provides immediate feedback, helping them identify areas for improvement and deepen their appreciation of cultural nuances.

The Cultural Scribe offers a high level of interactivity through its advanced bot, which tailors conversations to individual learning needs. For example, if a user is interested in discussing cinema, the bot can delve into various aspects such as the history of film in different countries, prominent filmmakers, and iconic movies. It can also explore the cultural significance of certain genres or cinematic movements, and even suggest relevant films for the user to watch. By providing information and quizzes related to the user's specific interests, the bot ensures a more engaging and personalized learning experience. This real-time practice not only helps users refine their skills but also builds their confidence in navigating various cultural settings.

Together, these components of ICLS create a comprehensive cultural educational experience that mirrors the structure and engagement of a serious game. By combining extensive reference materials, interactive quizzes, and personalized simulations, ICLS provides a robust framework for developing cultural intelligence. This game-based approach ensures that users can effectively apply their learning in real-world scenarios, enhancing their ability to interact successfully across different cultures.

\section{Architecture}
ICLS has been developed to offer a gamified learning experience that engages CultureVo users who are focused on exploring and understanding the diverse intercultural nuances around the world. At present, in July 2024, it is hosted as a web application with SSL certificates from Cloudfare. It has a layered architecture consisting of Frontend which provides the interactive experience to users. It has a multi-tier Backend which is hosted over cloud compute and uses open source Large Language Models (LLMs) to populate content over the Front end.

\subsection{Frontend}

The frontend of ICLS has been built using the following languages: HTML, CSS, and JavaScript. Bootstrap is utilized for UI components and styling, while Tailwind CSS provides additional UI components and styling. jQuery is used for DOM manipulation and AJAX requests, and AJAX handles asynchronous data loading and interactions. For displaying charts and graphs in the dashboard, Chart.js or Recharts are utilized. Animate.css/Framer Motion are employed for adding animations.

\subsection{Backend}
ICLS's  backend is divided into two parts: 1) Layer 1 consists of databases and web application. 2) Layer 2 consists of the Cultural Treasury, World Wise and Culture Scribe, and is an LLM powered application.

\subsubsection{Backend: Layer 1}
ICLS uses PHP and the Laravel PHP framework for building the web application. Laravel Socialite has been used for authentication with social networking platforms like Google and Facebook. OAuth 2.0, the industry-standard protocol, has been employed for authorization. Laravel Passport, an OAuth2 server and API authentication package, has been used for handling OAuth2 in Laravel. MySQL has been used to store user credentials, user progress data, and course content for CultureVo. It is a relational database management system installed on the hosting server. phpMyAdmin, a GUI-based application, has been used to administer this MySQL database. Eloquent, an object-relational mapper (ORM) and part of the Laravel framework, has been used to facilitate the handling of database records by representing data as objects. It provides a layer of abstraction on top of the database engine used to store the data.

\subsubsection{Backend: Layer 2}
Layer 2 of the backend includes the Cultural Treasury, World Wise, and the Cultural Fluency Trainer, all of which are powered by Generative AI. It helps generate summary and quiz for each video/document part of the ICLS database.

\subsubsection*{Cultural Treasury}
The Cultural Treasury serves as a comprehensive archive of reference materials, providing thorough insights into the cultural dimensions of various countries. This component features detailed glossaries, thematic summaries, and analytical frameworks that clarify the terminology and cultural themes explored in the lessons. The thematic summary generation feature is implemented using the LLM provided by Gorq API. The Gorq SDK is a powerful service for interacting with open-source LLM models like Llama 3 and Mistral. We are using Llama3 here as it has demonstrated to provide robust performance in RAG pipeline \cite{b2024evaluatingefficacyopensourcellms} \cite{juvekar2024introducingnewhyperparameterrag}. Admins can upload supported documents (.pdf, .txt, and .doc files) or a YouTube video link. First, text is extracted from the document or YouTube video. Then, a prompt is developed using the admins' instructions and the extracted text, which is passed to the Gorq API. If the given prompt and the extracted text fit within the model’s context window, the model will generate a summary, which is then returned via the API. The summary prompt given looks as following.

\subsubsection*{Summary Prompt:}
\noindent\textit{"Instruction: You are a summary generator, your job is to generate a summary of the given data.
You have to follow the instructions given on how to generate the summary. 
If no instruction is given,then just generate the summary. 
Data: \{combined text\} User Instruction: \{userPrompt\} \\
Instruction: The summary should be at least 200 words long \\ 
Summary:"}

\subsubsection*{World Wise}
World Wise incorporates an interactive quiz format that challenges users to assess the cultural knowledge they’ve acquired from the Cultural Treasury. The quiz generation feature involves feeding a file from the Treasury into the ICLS, which then generates a quiz based on the document. The quiz prompt given looks as following. \\

\subsubsection*{Quiz Prompt:}
\noindent\textit{"Generate a quiz based on the following information: Data: {combined text} "
            Instructions :\\
            1. Generate a quiz based on the given information.\\
            2. The quiz should be at least 10 questions long.\\
            3. The quiz should be in the form of a list of questions and options."
            Format of the quiz:\\
            Each question should be start with *Question :** \\
            Option should be start with *Option :** \\
            Each answer should be like *Answer :** and only give the option number for the answer \\
            Options: 1, 2, 3, 4
            Answer: Answer"}

\subsubsection*{Culture Scribe}
The Culture Scribe implements a chat with a video/ document feature by adding a Vector Store, where vector embeddings for the uploaded document or video transcript are created. Then, the most relevant text chunks are extracted from the Vector Store based on a cross between Cos-mix  and Keyword augmented retrieval approach \cite{juvekar2024cosmixcosinesimilaritydistance} \cite{purwar2023keywordaugmentedretrievalnovel}. Then, the retrieved  chunks are passed as input to the LLM model as a context with the user's question in the form of a prompt. Based on that, LLM responds to the question which is returned as answer to the user. This looks similar to how VidyaRang \cite{harbola2024vidyarangconversationallearningbased} provides a  chat functionality for each course. In long run, we envisage to provide a conversational learning experience  by creating country specific companion personalities based on the learning material for each country. The personality shifting companion would talk over a voice interface with the user.

\subsubsection*{Chat Prompt:}
\noindent\textit{
Role: You are a Question Answer solver. Here is the information: \\ Data: \{text \} \\
Using this information, answer the following question: \\
Question: \{userPrompt\} \\
Instruction: Answer the question using information provided in data. \\
Answer::}

\begin{figure*}
    \centering
    \includegraphics[width=\textwidth]{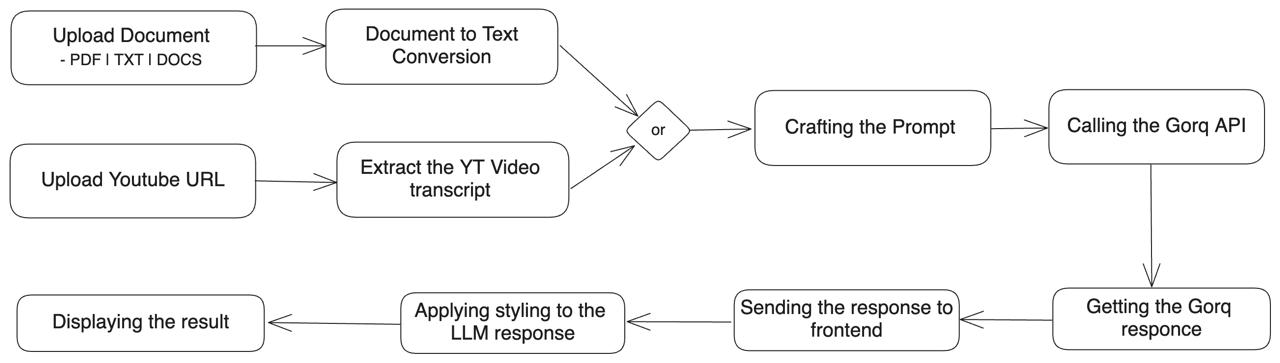}
    \caption{ICLS: Backend Layer-2}
    \label{fig:backflow2}
\end{figure*}

\subsubsection{DevOps and Deployment}
Apache Web Server has been used to deploy and serve the webpages of CultureVo. The Apache server has been hosted over EC2 instance offered by AWS. EC2 instance has been mapped to the web domain using GoDaddy and Cloudfare.


\section{Demo}
\begin{figure*}
    \centering
    \includegraphics[width=0.92\textwidth]{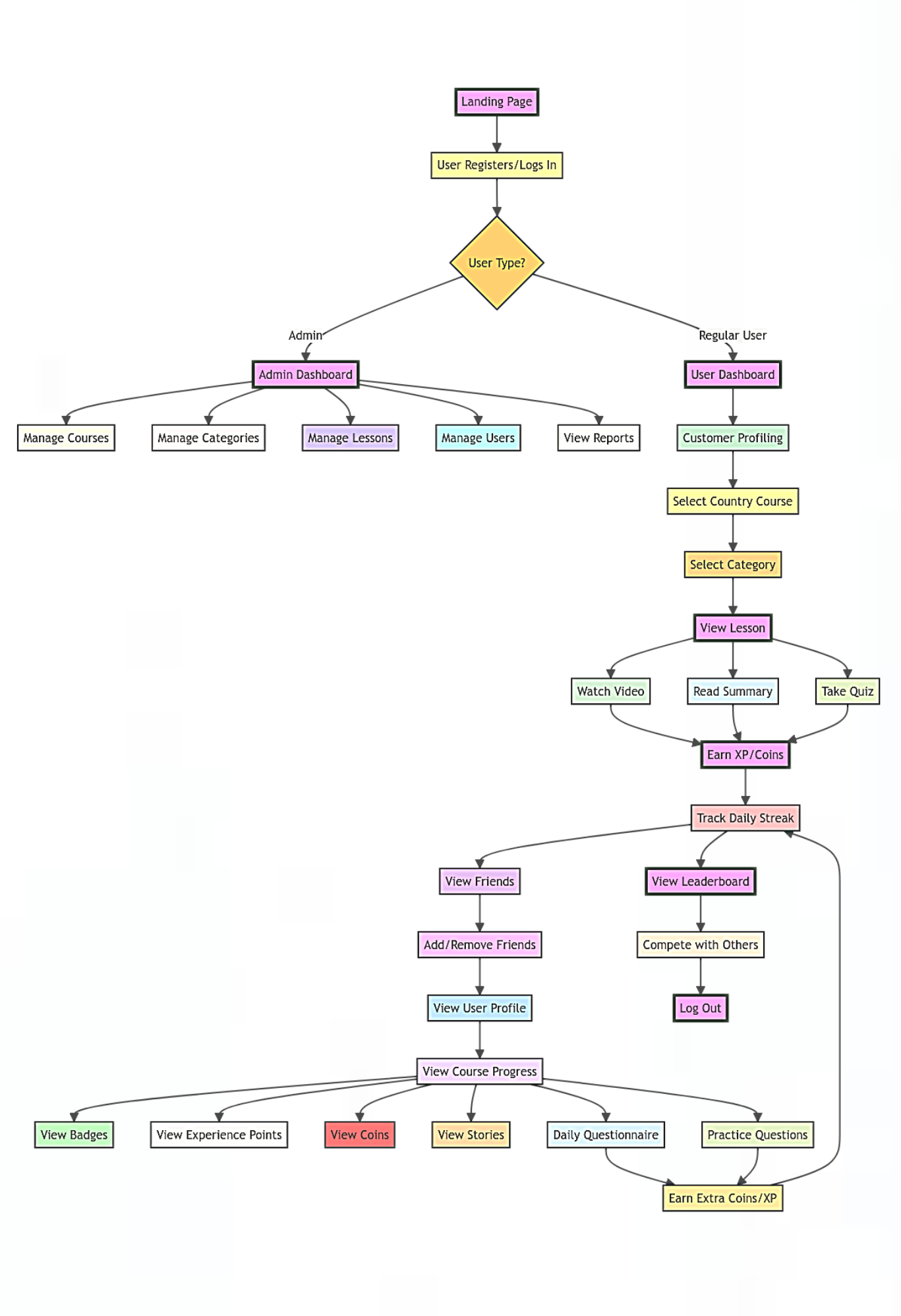}
    \caption{ICLS: Detailed Process Flowchart}
    \label{fig:detflow}
\end{figure*}

The ICLS framework has been implemented for \href{https://culturevo.com/} {CultureVo here}. The user needs to follow these steps once they register as a New User or Login. A first time user need to register as new users and then login as existing users, as shown in flowchart Figure \ref{fig:detflow}.
\begin{itemize}
    \item Immerse in a country: Users can enroll in the culture courses of different countries as available in the Cultural Treasury.
    \item Explore Categories: Each country's culture course contains categories like Art, Music, Cinema, Literature, Festivals, Fashion, Cuisine, Beverage, Customs, Dance, and Travel.
    \item Complete Lessons: Each category contains lessons with videos, summaries, and quizzes.
    \item Earn Rewards: Users earn coins, XP, badges and maintain daily streaks by completing lessons.
    \item Social Features: Based on the badges, the users can add friends and view leaderboard to compete with others.
    \item Logout: Users can log out when they have finished using the application.
\end{itemize}

\subsection{Landing Page:}
The landing page provides an inviting and
  visually appealing entry point for users. It has two Buttons:
 
\begin{itemize}
  \item
    \textbf{Get Started:} Redirects to the Sign-Up Page where users can
    enter details about their learning goals and preferences.
  \item
    \textbf{Already Have an Account:} Redirects to the Login Page for
    existing users.
  \end{itemize}

\subsection{Sign-Up and Login Flow:}
A user can register using Sign-Up and create a profile by entering following details. Once the user completes the registration, he/she is redirected to the User Dashboard.
  \begin{itemize}
  \item
    \textbf{Personal Details:}  Name and email address
    
    \item
      \textbf{Country Immersion:} Select a country of interest.
    \item
      \textbf{Learning Motivation:} Describe the reason for learning
      about the selected country\textquotesingle s culture.
    \item
      \textbf{Current Knowledge:} Rate current knowledge of the
      country's culture.
    \item
      \textbf{Daily Learning Goal:} Set a daily learning target.
    \item
      \textbf{Notification Preferences:} Opt-in for daily reminders and
      cultural notifications.
    \end{itemize}

\subsection{Login Page and Dashboard:}
User needs to perform Authentication using Google Sign-In or email and password. Once logged in, user is taken to the dashboard page which has following features.

  \begin{itemize}
  \item
    \textbf{Navigation Bar (Left Sidebar):}

    \begin{itemize}
    \item
      \textbf{Learn:} Access culture categories and lessons inside them.
    \item
      \textbf{Explore:} Select other countries to immerse in.
    \item
      \textbf{Practice:} Engage with practice questions.
    \item
      \textbf{Stories:} Listen to podcasts which have stories of intercultural experiences.
    \item
      \textbf{Leaderboard:} View rankings based on XP and compare with others.
    \item
      \textbf{Profile:} Access you user profile.
    \item
      \textbf{Account Settings:} Modify account details.
    \item
      \textbf{Logout:} Exit the application.
    \end{itemize}
  \item
    \textbf{Main Dashboard (Center Content):}

    \begin{itemize}
    \item
      \textbf{Country Name:} Displays the country currently being
      explored.
    \item
      \textbf{Categories:} Shows various categories with images, lesson
      count, progress and a view button.
    \item
      \textbf{Journey So Far:} Overview of course progress of the country currently being explored.
    \item
      \textbf{Achievements Box:} Displays latest achievements, badges,
      and milestones.
    \end{itemize}
  \end{itemize}

\subsection{Category Interaction:}
ICLS framework prioritizes on  user interaction by providing easy to use interaction points.
\begin{itemize}
\item
  \textbf{Lesson Cards:} Displays lesson images, names, number of
  videos, and a start button.
\item
  \textbf{Video Cards:} Displays Video images, names, progress and watch
  button.
\item
  \textbf{Video Content:}

\begin{itemize}

\item \textbf{Watch Video:} Engage with video content.

\item \textbf{Read Summary:} Brief summary and key points from the
video.

\item   \textbf{Take Quiz:} Answer questions related to the lesson,
receive feedback, and earn XP/Coins.

\item  \textbf{Quiz Navigation:} Options for previous, next, and skip.
Question indicators (current: orange, answered: green, unanswered:
silver, wrong: red).

\item  \textbf{Chatbot Interaction:} Access the chatbot for assistance
related to lesson content.
\end{itemize}
\end{itemize}
\subsection{User Engagement and Progression:}
To ensure an engaging learning experience, the ICLS framework utilizes the Gamification Octalysis Framework \cite{chou2019actionable} and incorporates the following elements of gamification:
\begin{itemize}
\item 
    \textbf{Experience Points (XP):}

    \begin{itemize}
    \item
      \textbf{x5:} Awarded for watching videos.
    \item
      \textbf{x7:} Awarded for watching videos and completing
      summary tests.
    \item
      \textbf{x12:} Awarded for watching videos, completing summary
      tests, and practice tests based on the interaction with the Culture Scribe bot.
    \end{itemize}
  \item
    \textbf{Coins:} Earned from daily challenges and quizzes.
  \item
    \textbf{Badges:} Awarded for completing full immersion in a country's course (country badge) and also for achieving
    mastery in specific categories (e.g., History Badge,
    Art Badge).

  \item
    \textbf{Track Streak:} Monitor daily logins for a user.
  \item
    \textbf{Daily Challenges:} Engage with daily quizzes to coins.

\item
  \textbf{Leaderboard Rankings:} Compare XP with others globally, within the
    country, among friends, and within organizations.
\end{itemize}

\subsection{Social and Interactive Features:}
For users to have a community of like minded individuals who can engage in culturally conscious conversations, ICLS framework provides following:
\begin{itemize}
\item
  \textbf{Friends Network:} Connect with friends, send requests and view
  their achievements.
\item
  \textbf{Recommendation System:} Suggests courses based on shared
  interests and activities.
\item
  \textbf{Stories and Podcasts:} Access podcasts where users talk about their intercultural experiences, for deeper insights.
\end{itemize}

\section{Future Scope of Work}
A new layer is envisioned for the ICLS: the Cultural Fluency Trainer. This advanced interactive bot aims to enhance cross-cultural interaction skills through adaptive algorithms that provide personalized experiences based on users' interests and learning needs. By simulating real-world cultural scenarios, role-playing, and decision-making, users will navigate diverse cultural contexts and gain deeper insights into intercultural similarities and differences. Immediate feedback and reflection will aid in understanding these nuances, while varied content and challenges will expose users to a broad range of cultural perspectives. The bot’s collaborative elements will promote cross-cultural communication and cooperation, and its adaptability will ensure personalized learning experiences.  In long run, we envisage to provide a conversational learning experience by creating country specific companion personalities. The
personality shifting companion would talk over a voice interface with the user. This comprehensive approach will help users refine their cultural competency and build confidence, enhancing their ability to communicate effectively across diverse environments.

\section{Conclusions}
CultureVo' ICLS framework has been built using a traditional 2 tier architecture consisting of frontend and backend hosted over cloud compute instance. Its backend uses Laravel (a php framework) and elements of Generative AI in the form of open source LLMs viz. Llama 3. Connecting the php backend and mysql database is the LLM based pipeline which generate  summary and quiz for each lesson based on tailored prompts. The technical stack of CultureVo's ICLS is designed to deliver the benefits of enhanced cultural intelligence for both individuals and organizations. For individuals, cultural intelligence fosters better interpersonal relationships, improves communication skills, and enhances adaptability in diverse environments, which can lead to more successful personal and professional interactions. For organizations, it facilitates smoother cross-cultural collaboration, enhances global business strategies, and helps in managing diverse teams more effectively. This capability can lead to increased innovation, improved customer relations, and a competitive edge in the global market, ultimately contributing to organizational success and growth. In essence, CultureVo will empower individuals and organizations to enhance their cultural intelligence, unlocking unprecedented opportunities and driving excellence in an increasingly interconnected world, while addressing the expanded responsibility that globalization and technology have placed on individuals for the well-being of people worldwide.

\bibliographystyle{ACM-Reference-Format}
\bibliography{bibliography}

\end{document}